# Photo-thermoelectric properties and their use in study of transport properties of both carriers from a single bulk sample

Zhenyu Pan[1], Zheng Zhu[1], Jeffrey J. Urban[2], Fan Yang[3], and Heng Wang[1]

1. Department of Mechanical, Materials, and Aerospace Engineering, Illinois Institute of Technology, Chicago, IL 60616, USA
2. The Molecular Foundry, Lawrence Berkeley National Laboratory, Berkeley, CA 94720, USA
3. Department of Mechanical Engineering, Stevens Institute of Technology, Hoboken, NJ 07030, USA

Abstract: We describe a theory on photo-thermoelectric properties of a semiconductor, which include photo-conductivity, photo-Seebeck coefficient, and photo-Hall effect. We demonstrate that these properties provide a powerful tool for the study of carrier transport in semiconductors. Even though photo-carrier generation is a complicated process which often prohibits quantitative analysis as their species or numbers are not known. Using bulk samples seems even less likely as the photo-carrier only affect a thin layer. Our method will allow researchers to bypass these difficulties, to use only measured properties and determine both electron and hole mobilities as well as the ratio between electrons and holes from a bulk sample. We provide initial experiment verification of our theory in the end using two distinctively different semiconductors.

Thermoelectric properties describe a semiconductors potential for use in thermoelectric devices, which are extensively studied for applications in power generation and solid state cooling[1]. Meanwhile, these properties are widely studied in semiconductor research for fundamental transport parameters. Taking Seebeck coefficient ($S$) as an example, $S$ often used to determine carrier type. Its temperature dependence can be used to determine bandgap $E_g$. The carrier density dependence of $S$ can be used[2] to determine carrier effective mass $m^*$, (Pisarenko relation[3]), while the $S$ vs $\ln\sigma$ relation[4] (first reported by G. Jonker in 1968) provides information on the product $\mu_0 m^{*3/2}$ ($\mu_0$ is mobility at low doping levels). Plenty of examples exist in literature on thermoelectric material research[5-9], thus will not be elaborated here.

When a semiconductor is under photo illumination, its thermoelectric properties will be different from conventional ones because photons generates photo-carriers. While thermoelectric properties are being studied, and used widely, photo-thermoelectric properties of a semiconductor have received little attention.

Based on transport theory, we will show in this work that combining three properties: photo-conductivity, photo-Seebeck effect, and photo-Hall effect, (which we call photo-thermoelectric properties because they are properties often measured together in thermoelectric research, even though conductivity and Hall effect are not thermo-electric properties by definition), could allow researchers to resolve a lot more information regarding the photo-carriers and their transport: mobilities of electrons and holes, densities of each of them (or their ratio), even effective masses of each carriers. Using regular approach, these parameters will require multiple samples with different dopant concentrations being made and evaluated. With photo-thermoelectric measurements, they can be determined from a single, intrinsic sample, even when a certain type of doping is unachievable. Carrier mobilities are key parameters for essentially all semiconductor applications, whereas being able to determine carrier densities and their mobilities individually as a result of certain excitation condition could greatly help the understanding of defects in materials. Thus, we believe this work is of general interest to researchers across different areas.

Figure 1 below illustrates how each property is determined. They can be seen as analogs of standard transport properties, evaluated under illumination. Take photo-Seebeck effect as an example: a homogeneous semiconductor is illuminated by a uniform, continuous light. A steady temperature gradient is applied perpendicular to the illumination. An open circuit voltage is observed across the semiconductor along this direction, which is proportional to the temperature difference. The term "photothermoelectric effect" is often used in literature for a related yet fundamentally different phenomenon: the detection of photocurrent under localized illumination on devices made of graphene or other atomically thin materials. This is not to be confused with effects discussed here.

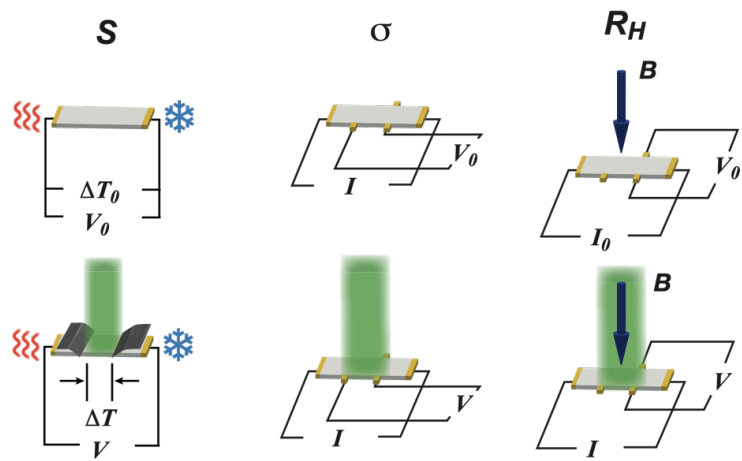

Fig. 1. Illustrations of thermoelectric and photo-thermoelectric measurements. Note masks are used for photo-Seebeck measurement to remove photovoltaic effect at contacts. Measured quantity (V-$V_0$)/$\Delta T$ reflects the difference between photo-Seebeck coefficient and regular Seebeck coefficient at dark.

Both photo-Seebeck and photo-Hall effect are known but has not received much attention. In the 1970s R. Bube and his group studied Seebeck coefficients of bulk GaAs and Si, as well as CdS films under photon illumination[10, 11]. Later Terazaki and his group reported Seebeck coefficients in bulk ZnO[12] PbO[13], and[14] $PbCr_2O_5$ illuminated with photon of different wavelengths. As important as these works are, they lack adequate analysis and interpretation of results, no practical application was demonstrated either. Similar history is found for photo-Hall effect, where[15, 16] it was studied and analyzed based on a two-carrier model.

In a very recent high-profile publication[17]. Photo-Hall effect was used to study carrier mobilities of both types. As will be shown here this can be seen as a special case when excited electrons and holes are equal. In this case, photo-Hall effect is very convenient offering carrier-specific mobilities and densities with only one measurement. Nonetheless, we present here a more general, versatile tool for materials research.

We derived the analytical relations between Seebeck coefficient ($S$), conductivity $\sigma$ and Hall coefficient ($R_H$) under illumination. Fig.2 shows such relations simulated for a simple case in a film, with equal number of photo-generated electrons and holes. Material parameters used for this simulation are: $m_e^* = m_h^* = 0.3$, intrinsic carrier density $n_{h,0} = 1\times10^{12}$ cm$^{-3}$. Blue dots are for different cases with different values on $\beta = |\mu_e/\mu_h|$. The green line indicates regular chemical doping (assumed n type) behavior.

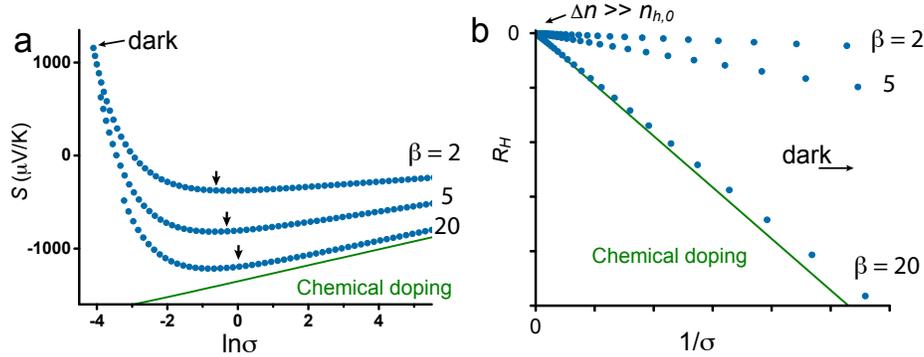

Fig. 2. Simulated relations a) $S$ and $\ln\sigma$ (arbitrary unit), b) $R_H$ and $1/\sigma$ (both in arbitrary units), for both conventional chemical doping (n-type) and photon doping, assuming different $\beta$ values.

Before discussing the theory in detail, we should consider the complex nature of photo-carriers.

Photons excite carriers in semiconductors, which we assume, for most inorganic semiconductors at room temperature and above, are free carriers. The exact result of photon excitation (or say, photon doping), varies under different conditions: As most commonly regarded, photons above the bandgap cause band-to-band excitations, creating electron-hole pairs in equal numbers ((1) in Fig.3). The initial relaxation towards band edges can be considered instant, so the excited carriers populate lowest allowed energies from band edge. The electrons will stay for some time before they recombine with holes. This time scale is the (minority) carrier lifetime, which is long (>10$^{-6}$ S) in high quality, photo-sensitive semiconductors[18, 19]. For instance, in Si minority electrons can live up to[20] 1 mS. Under continuous illumination, there is a steady number of photon-generated electrons and holes, which number is proportional to minority carrier lifetime and illumination intensity.

This picture works well for some materials, but in general is over-simplified because in-gap defect levels could play a critical role. There are two categories of defects, the first is shallow defect, whose energy difference from band edge is small. Its impact can be viewed as reducing mobility without reducing the number of carriers, as thermal reactivation will happen. More importantly is the second type -- deep defects. Carriers trapped by deep defects are removed from transport. If the defects trap electrons and holes at similar rates, they are called recombination centers ((1) in Fig.3). Compare this with defect-free conditions, only the lifetime is reduced, the numbers of electrons and holes are still equal. In other cases, deep defects could capture primarily one type of carrier, causing an imbalance[18, 21] between free electrons and holes ((2) and (3) in Fig.3). This is often seen in compound semiconductors with large numbers of deep levels. Interestingly, this does not necessarily mean a decrease of photo-carrier densities or damp of a materials photo-response, instead, this is a common strategy used in early development of photo-conductors to enhance their sensitivity (called sensitization[18, 22, 23]). As one type of carriers are selectively trapped by defects the other type of carriers are protected from recombination, leading to multiple-orders-of-magnitude increase of their lifetime. In addition, excitation could happen between bands and defects levels. Sub-bandgap photo-response is commonly seen in semiconductors[24-27] (often not as strong). This could also create unequal numbers of free electrons and holes. First, this can be seen in (6) and (7) of Fig.3, where the photon energy is only enough to excite carriers to/from only one side of the gap. In extreme cases this will only create one type of carriers, although in reality, the defects' ability to trap different carriers will change in response, potentially leading to notable numbers of the opposite carriers as well. Second, photons with energy enough to excite carriers to/from both sides of the gap, the electron and hole densities can also be unequal, depending on the defects' position. This is because, defect levels are very narrow (effectively at a single energy level). How often a transition happens is related to the density of available states on both the initial and final states. The filled/open states of the defect levels can vary, but the dominant factor should be the density-of-state in the bands at a specific energy (to allow transition), which increase very rapidly ($DOS \propto E^{1/2}$) as it moves into the band. So, if the defect is closer to the conduction band, there will be more electrons excited. The opposite is true for holes, as illustrated in (4) and (5) in Fig.3.

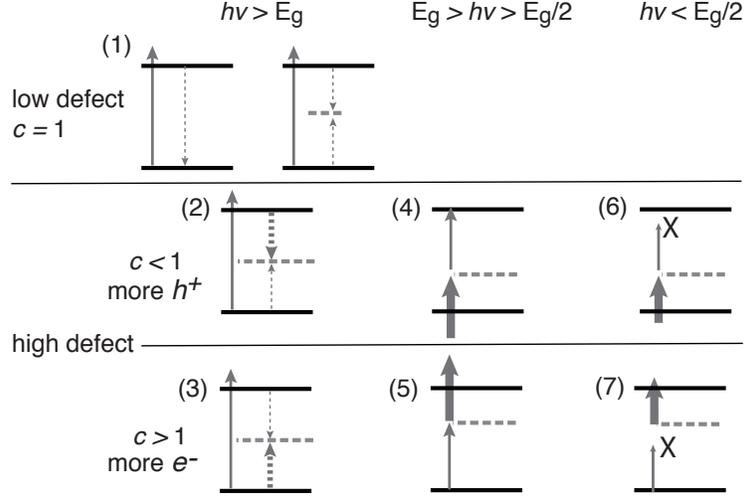

Fig. 3. Different excitation conditions and defect structures could lead to different electron and hole density ratios, $c$.

As a general conclusion, photons lead to free carriers for transport, but not necessarily in equal numbers. This can be described by a ratio $c$ between their densities, $c = \Delta n_e/\Delta n_h$.

We then apply standard transport theory to the transport properties with photo-carriers, this is valid because classic transport theory applies to photo-carriers. This is further because: 1) transport and diffusion of photo-carriers are perpendicular to each other thus only transport terms exist along the measurement direction. 2) Fermi distribution applies because hot carrier cooling is very fast. 3) The transport processes reflect bulk properties in thick samples due to photo-carrier diffusion.

The approximations we made are: 1) photo-carrier density profile due to diffusion towards inside of the material is approximated with an effective phto-carrier layer with uniform carrier distribution, for thin film samples the photo-carriers populate the entire sample, whereas for bulk samples there is a top photo-carrier layer and a bottom dark layer, each have uniform properties. This assumption is commonly used[12, 16] in study photo-carrier transport. 2) carrier mobilities doesn't change with carrier density. constant mobilities insensitive to carrier density is a commonly observed behavior in many semiconductors[20] like Si or GaAs.

Let's consider a p-type thin film with intrinsic carrier density $n_{h,0}$, and write $b = n_{h,0}/\Delta n_h$. $\Delta n_h$ is photo-generated free hole density, correspondingly the free electron density will be $\Delta n_e = c\Delta n_h$.

The conductivity can be written as:

$$\sigma = en_e\mu_{e,0} + en_h\mu_{h,0} = e\mu_{h,0}\Delta n_h[(b+1) + c\beta] \qquad \text{eq.1}$$

The two-carrier Seebeck coefficient:

$$S = \frac{S_e n_e e\mu_{e,0} + S_h n_h e\mu_{h,0}}{n_e e\mu_{e,0} + n_h e\mu_{h,0}} = \frac{(b+1)S_h - c\beta S_e}{(b+1)+c\beta} \qquad \text{eq.2}$$

$S_e$, $S_h$ are the Seebeck coefficients of electrons and holes. $\beta = |\mu_{e,0}/\mu_{h,0}|$, $\mu_{e,0}$ and $\mu_{h,0}$

are constant electron and hole mobilities at low doping levels. $S_e$ and $S_h$ are determined via:

$$|S_{e(h)}| = \frac{k_B}{e}\left(\frac{5}{2} + r + \ln\frac{2(2\pi m^*_{e(h)}k_BT)^{3/2}}{h^3 n}\right) \quad \text{eq.3}$$

$r$ is a constant determined by the dominant carrier scattering mechanism. Substituting eq. 3 into eq. 2:

$$S\left(\frac{k_B}{e}\right)^{-1} = \frac{c\beta-(b+1)}{c\beta+(b+1)}\ln\Delta n_h + \left\{\frac{(b+1)-c\beta}{c\beta+(b+1)}\left(\frac{5}{2}+r+\ln\frac{2(2\pi k_BT)^{3/2}}{h^3}\right)+\right.$$

$$\left.\frac{3}{2}\frac{(b+1)\ln m^*_h - c\beta\ln m^*_e}{c\beta+(b+1)} + \frac{c\beta\ln c - (b+1)\ln(b+1)}{c\beta+(b+1)}\right\} \quad \text{eq.4}$$

Using eq. 1 to substitute $\Delta n_h$ with $\sigma$, and write $f = c/(b+1)$:

$$S\left(\frac{k_B}{e}\right)^{-1} = \frac{f\beta-1}{f\beta+1}\ln\sigma + \left\{\frac{1-f\beta}{1+f\beta}\left(\frac{5}{2}+r+\ln\frac{2e(2\pi k_BT)^{3/2}}{h^3}+\ln\mu_{h,0}\right)+\right.$$

$$\left.\frac{3}{2}\frac{\ln m^*_h - f\beta\ln m^*_e}{1+f\beta} + \frac{f\beta\ln f + (1-f\beta)\ln(1+f\beta)}{1+f\beta}\right\} \quad \text{eq.5}$$

Eq.5 is the general equation describing the relation between measured $S$ and $\sigma$ under the same illumination condition.

Similarly, consider two-carrier Hall coefficient:

$$R_H = \frac{r_H}{e}\frac{n_e\mu^2_{e,0}-n_h\mu^2_{h,0}}{(n_e\mu_{e,0}+n_h\mu_{h,0})^2} \quad \text{eq.6}$$

The Hall factor $r_H = 1.18$ for low carrier density conditions. we can get:

$$R_H = 1.18\mu_{h,0}\frac{1-f\beta^2}{1+f\beta}\frac{1}{\sigma} \quad \text{eq.7}$$

In general cases, we can only solve for $f$ and $\beta$ by combining $S$, $R_H$ and $\sigma$.

Eq. 5 and 7 shall be discussed under different conditions:

1) $c \approx 0$. This represents an extreme case where only free holes are created by sub-bandgap photons ((6) in Fig. 3), or all excited minority electrons are trapped by deep defects ((2) in Fig.3). In this case $f \approx 0$, eq. 5 and 7 became:

$$S\left(\frac{k_B}{e}\right)^{-1} = -\ln\sigma + \left[\left(\frac{5}{2}+r+\ln\frac{2e(2\pi k_BT)^{3/2}}{h^3}\right)+\ln\mu_{h,0}m^{*3/2}_h\right] \quad \text{eq.8}$$

$$R_H = 1.18\mu_{h,0}\frac{1}{\sigma} \quad \text{eq.9}$$

These are the same as commonly used relations for standard thermoelectric properties. $S$ vs $\ln\sigma$ is the Jonker relation, its slope is a physical constant $-k_B/e$, its intercept can be used to calculate the product $\mu_{h,0} m_h^{*3/2}$. With $\mu_{h,0}$ determined from $R_H$, $m_h^*$ can be found. $\Delta n_h$ can also be determined from $\sigma$.

2) $c > 0$, $b \ll 1$. This is the case where intrinsic carrier density is low, light intensity is high. Eq. 5 and 7 now becomes:

$$S\left(\frac{k_B}{e}\right)^{-1} = \frac{c\beta-1}{c\beta+1}\ln\sigma + \left[\frac{1-c\beta}{1+c\beta}\left(\frac{5}{2}+r+\ln\frac{2e(2\pi k_B T)^{3/2}}{h^3}+\ln\mu_{h,0}+\ln(1+c\beta)\right)+\frac{3}{2}\frac{\ln m_h^*-c\beta\ln m_e^*}{1+c\beta}\right] \quad \text{eq.10}$$

$$R_H = 1.18\mu_{h,0}\frac{1-c\beta^2}{1+c\beta}\frac{1}{\sigma} \quad \text{eq.11}$$

For the special case with $c = 1$, Both $S$ and $R_H$ alone can be used to determine $\beta$. Individual mobilities can then be calculated with $\mu_{h,0}$ from dark Hall effect. For more general cases however, $\beta$ and $c$ can only be solved by combining eq.10 and 11. $\Delta n_h$ and $\Delta n_e$ can be determined from $\sigma$. For the effective masses, $m_h^*$ can be obtained from dark $S$ and $n_H$ (Pisarenko relation), then $m_e^*$ can be solved from the intercept of eq.10. The effective masses are determined from single measurement point or intercept, which could potentially lead to large uncertainties.

3) $c \gg 1$, $b \ll 1$. This is the other extreme case where only free electrons are created ((7) in Fig.3), or most holes are trapped ((3) in Fig.3). Here $f \gg 1$, eq. 5 and 7 become:

$$S\left(\frac{k_B}{e}\right)^{-1} = \ln\sigma + \left[\left(\frac{5}{2}+r+\ln\frac{2e(2\pi k_B T)^{3/2}}{h^3}\right)+\ln\mu_{e,0}m_e^{*3/2}\right] \quad \text{eq.12}$$

$$R_H = 1.18\mu_{e,0}\frac{1}{\sigma} \quad \text{eq.13}$$

$S$ vs $\ln\sigma$ has a fixed slope $k_B/e$. $R_H$ vs $1/\sigma$ measures electron mobility. The intercept of $S$ vs $\ln\sigma$ can be used to calculate $\mu_{e,0} m_e^{*3/2}$.

In cases 2) and 3), we assumed $b \ll 1$, the photo-carrier density has to be much greater than the intrinsic one. For general conditions where this is not met, analytical relations can't be found, numerical fitting based on full equations is needed. As seen in Fig. 2, the linearity of $R_H$ vs $1/\sigma$ hold for a wider range, which is because of the direct dependence of $R_H$ on $n$.

Now consider a thick, bulk sample, which contains a top, photo-carrier populated layer (thickness $d$) and an underlying, dark layer (thickness $D$). Observed properties will be combined contributions from these two layers. Usually to quantitatively study the change brought by photon-generated carriers, one needs to know: $d$, $\Delta n_h$ and $\Delta n_e$, etc. These further depend on details of absorption and diffusion processes, which are difficult to evaluate. Nonetheless, we suggest that this can be bypassed by analyzing photo-thermoelectric properties directly.

Write $S$ ($R_H$, $\sigma$), $S_{ph}$ ($R_{H,ph}$, $\sigma_{ph}$), and $S_0$ ($R_{H,0}$, $\sigma_0$) as the measured properties under illumination, the properties of the top layer, and the dark layer, respectively. $\sigma_{ph}$, $S_{ph}$, and $R_{H,ph}$ are expressed by two-carrier transport equations eq. 1, 2 and 6. The measured properties are averages written as:

$$\sigma = \frac{\sigma_{ph}d + \sigma_0 D}{d+D} \qquad \text{eq. 14}$$

$$S = \frac{S_{ph}\sigma_{ph}d + S_0\sigma_0 D}{\sigma_{ph}d + \sigma_0 D} \qquad \text{eq. 15}$$

$$V_H = \frac{V_{H,ph}\sigma_{ph}d + V_{H,0}\sigma_0 D}{\sigma_{ph}d + \sigma_0 D} \qquad \text{eq. 16}$$

$V_H$ is the Hall voltage, which is what measured in Hall effect measurements and gives Hall coefficient $R_H$ via $R_H = V_H t/IB$, $t$ is the layer thickness, $I$ is the current passed through that layer, and $B$ is the magnetic field strength. The partition of total current between two layers is based on their conductance, so eq.16 lead to:

$$R_H = \frac{R_{H,ph}\sigma_{ph}^2 d + R_{H,0}\sigma_0^2 D}{\sigma_{ph}d + \sigma_0 D} \frac{d+D}{\sigma_{ph}d + \sigma_0 D} \qquad \text{eq.17}$$

If the photo-response is strong such that $\sigma_{ph} - \sigma_0 \approx \sigma_{ph}$, so that:

$$\sigma - \sigma_0 = \frac{d}{d+D}(\sigma_{ph} - \sigma_0) \approx \frac{d}{d+D}\sigma_{ph} \qquad \text{eq.18}$$

based on eq. 14, 15 and 18 we find:

$$S - S_0 = \frac{\sigma_{ph}d}{\sigma_{ph}d + \sigma_0 D}(S_{ph} - S_0) = \frac{\sigma - \sigma_0}{\sigma}(S_{ph} - S_0) \qquad \text{eq. 19}$$

We can now use measurable quantities to describe $S_{ph}$, which is the measured Seebeck coefficient in the previous thin film case. However, $\sigma_{ph}$ can't be solved, for this we assume a general relation[18] between incident light intensity $I_{ph}$ and $\sigma_{ph}$: $\sigma_{ph} = \phi I_{ph}^\alpha$, where $\phi$ and $\alpha$ are constants. We find:

$$(S - S_0)\frac{\sigma}{\sigma - \sigma_0}\left(\frac{k_B}{e}\right)^{-1} = a\frac{f\beta - 1}{f\beta + 1}\ln I_{ph} + \left\{\frac{1-f\beta}{1+f\beta}\left(\frac{5}{2} + r + \ln\frac{2e(2\pi k_B T)^{3/2}}{h^3} + \ln\mu_{h,0} - \ln\phi\right) + \frac{3}{2}\frac{\ln m_h^* - f\beta \ln m_e^*}{1+f\beta} + \frac{f\beta \ln f + (1-f\beta)\ln(1+f\beta)}{1+f\beta}\right\} - S_0\left(\frac{k_B}{e}\right)^{-1} \qquad \text{eq. 20}$$

where the exponent $\alpha$ can be determined via the slope from following equation:

$$\ln \sigma_{ph} = \ln\left(\frac{D+d}{d}\sigma - \frac{D}{d}\sigma_0\right) = \ln\frac{D+d}{d} + \ln\left(\sigma - \frac{D}{d+D}\sigma_0\right)$$

$$\approx \ln\frac{D+d}{d} + \ln(\sigma - \sigma_0) = a\ln I_{ph} + \ln\phi \qquad \text{eq.21}$$

Meanwhile for $R_H$, based on eq. 6 and 17, we find:

$$R_H = 1.18\mu_{h,0}\left[\frac{1-f\beta^2}{1+f\beta}\sigma + \frac{f\beta(1+\beta)}{1+f\beta}\frac{\sigma_0 D}{d+D}\right]\frac{1}{\sigma^2} \qquad \text{eq.22}$$

Eq.20 and 22 are the master equations for bulk case comparable to equation 5 and 7. Similar discussion on the three conditions apply for bulk as well. The major differences are: since $\sigma_{ph}$ is unknown, $\Delta n_h$ and $\Delta n_e$ can't be determined (their ratio can still be determined). Also, since $\phi$ is unknown, the effective masses, which were obtained from the intercept of $S$ vs $\ln\sigma$, will reflect their upper bounds.

The analysis is based on p-type sample, for n-type samples, similar analysis and conclusions can be found as well.

Here we provide some initial experimental verification of our theory: The first example is a piece of commercial low-doped, n-type Si wafer. It's conductivity at dark is $\sigma_0$ =0.025S/m, this correspond to a carrier density on the order of $10^{12}$ cm$^{-3}$. Small bar samples are cut off the wafer and used as-is. Illumination is from 565nm and 780nm LEDs. Si is an example of high-quality, defect-free semiconductors. Photo-response in Si is fast thus the Seebeck coefficient is measured using a AC technique with steady temperature gradients and light modulation from a mechanical chopper (1 Hz). Experimental relation between $(S-S_0)\sigma/(\sigma-\sigma_0)$ vs $\ln I_{ph}$ was found linear in both cases.

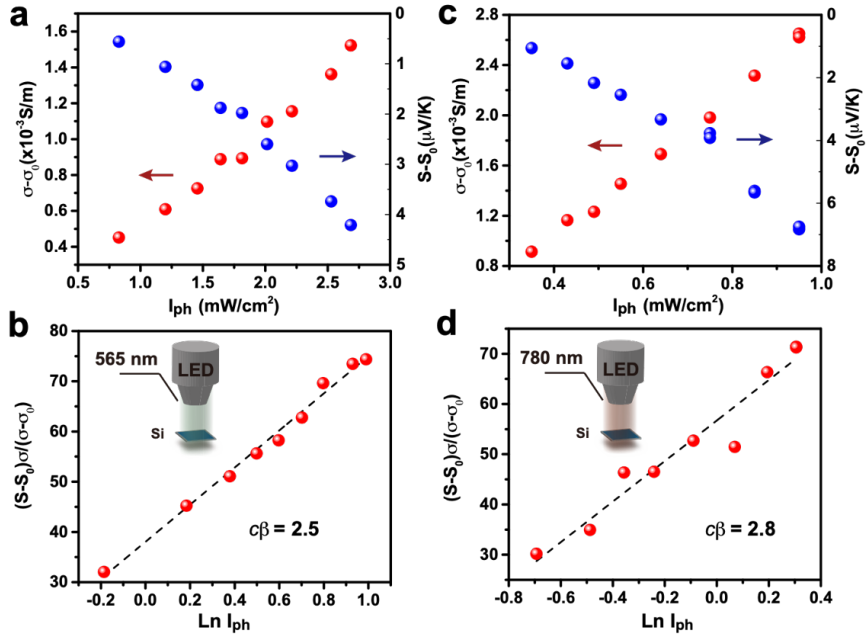

Fig. 4. Initial method validation using Si with 565 nm and 780 nm LED light sources. a), c) change of $\sigma$ and $S$ as a function of light intensity $I_{ph}$. b), d) $(S-S_0)\sigma/(\sigma-\sigma_0)$ vs $\ln I_{ph}$ showing a linear relationship, from the slope $c\beta$ is calculated to be 2.5 and 2.8, where the reference is 3. Unit for $I_{ph}$ is mW/cm$^2$

When bulk samples are used, the property changes in top photo-carrier layer are averaged by a much thicker, unilluminated background, which will significantly reduce the observable difference through measurements. In this experiment shown in Fig. 10, the change of Seebeck coefficient was no more than 4 μV/K (out of the dark value $S_0$ = -1700 μV/K) with 565 nm light source, and 7 μV/K with 780 nm source, respectively. We are able to detect such subtle trends because of the sensitivity provided by AC Seebeck measurement technique.

The slope indicated that photo-carriers are mixture of electrons and holes, the product $c\beta$ is found to be 2.5 and 2.8 with the use of 565nm and 780nm LEDs ($c$ is the ratio between photon-generated electrons and holes, in this case should be close to 1). Literature value[20] of $\beta$ for Si is around 3 (1500 cm$^2$/Vs for electrons and 500 cm$^2$/Vs for holes). Thus, our result provides a fairly accurate estimate of the mobility ratio $\beta$. When $\mu_e$ is determined from $R_H$, the minority carrier mobility can be determined from this n-type sample. Since both 780 nm and 565 nm photons are above Si bandgap, they create the same photo-carrier configuration. Small differences seen in specific values are likely due to the difference in LED spectral distribution, the exact number of photons, the difference in absorption coefficient thus different effective layer thickness, etc.

The second example is a piece of polycrystalline Se, which is obtained by melting 99.999% elemental Se in an evacuated quartz ampoule, then slowly cooled down to room temperature over 24 hrs. The sample was cut from ingot and polished to 0.3 mm thick. Se is a semiconductor crystalize in a hexagonal lattice with band gap around[28-30] 1.8 eV. It is p-type, with carrier density between[15, 31] 10$^{11}$ and 10$^{13}$ cm$^{-3}$.

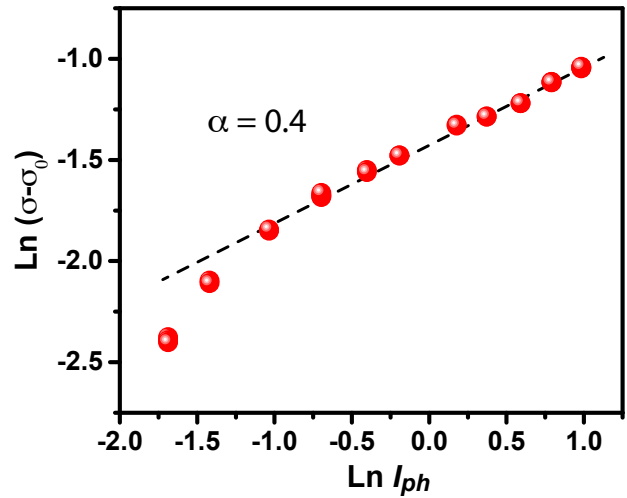

Fig. 4. The variation of observed photoconductivity as a function of photon intensity, the photoconductivity follows a power law relation with $I_{ph}$. indicating significant defect influence.

The detailed photo-response process in Se is very different from Si. A power-law dependence of $\sigma - \sigma_0$ on $I_{ph}$ is found. This is a signature of defects influence. Besides, Se shows persistent photoconductivity[32-34] over thousands of seconds after illumination is turned off whereas in defect-free Si this response is instant. This is a case with significant defect contribution and as a result the photo-carriers are likely of unequal number of electrons and holes.

Slow photo-response has prevented us from using the same AC technique to measure Seebeck coefficient. Instead, a AC measurement was performed with steady illumination under temperature modulation. The sensitivity will not be as high as with light modulation, but likely be more accurate than DC methods considering the sample resistance which is over MΩ. And fortunately, the change in Seebeck coefficient is significant enough to be observed ($S_0$ = +1250 µV/K). As a success of our theory, a linear trend is found between $(S-S_0)\sigma/(\sigma-\sigma_0)$ vs $\ln I_{ph}$, despite of the significant defect influence. From the slope, we know photo-carriers are also mixture of electrons and holes. The product $c\beta$ is found to be 0.6. Due to the strong defect influence, $c$ in this case is likely notably different from 1. Determination of individual $c$ and $\beta$ value is possible with $R_H$ measured. We are currently working a AC Hall effect measurement setup to allow us determine small changes more accurately.

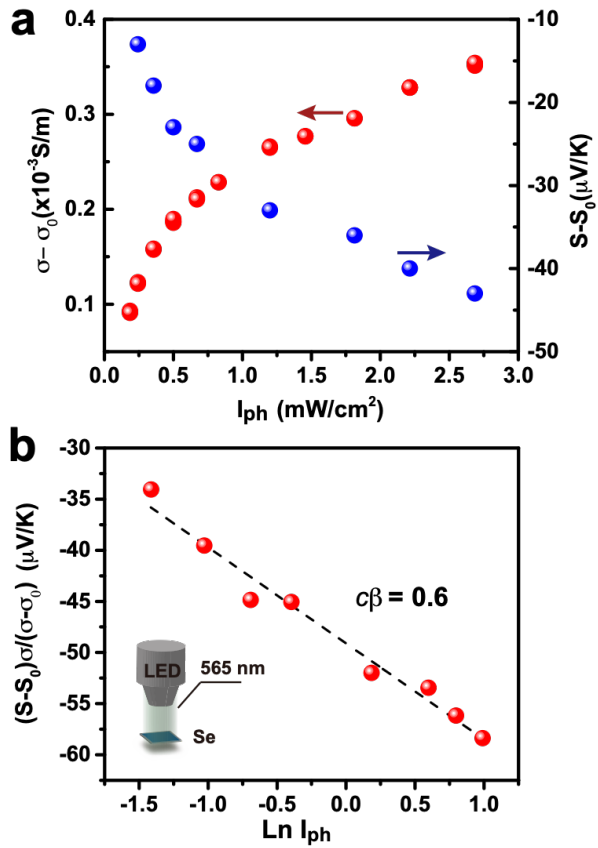

Fig. 5. Initial validation using polycrystalline Se. a) change of $\sigma$ and $S$ as a function of light intensity $I_{ph}$. b) $(S-S_0)\sigma/(\sigma-\sigma_0)$ vs $\ln I_{ph}$ showing a linear relationship, from the slope $c\beta$ is calculated to be 0.6. Unit for $I_{ph}$ is mW/cm$^2$

To conclude, we have derived the theoretical relation between photo-thermoelectric properties, specifically photo-conductivity, photo-Seebeck coefficient, and photo-Hall effect. We found the combined measurement is a powerful tool to study carrier transport in semiconductors. It allows the determination of mobilities, carrier densities, even effective masses of both electrons and holes, all from a single sample. With specially developed measurement technique, even bulk samples can be used. Compared with conventional transport study methods, or technique used to study photo-carrier transport, photo-thermoelectric measurement has great advantage: 1) it uses only a single sample, thin film or bulk samples can both be measured without altering the setup. 2) it provides carrier-type resolution, so individual properties of electrons and holes can be determined, whereas most other methods measure their combined influence. 3) it provides carrier density-mobility resolution. Photo-response of a material is otherwise observed through photo-conductivity or other property change due to conductivity change, with our method we can determine whether such response is due to changes of carrier densities, or carrier mobilities, thus offering extra insight into photo-carrier dynamics.